# Massive 1D Dirac Line, Solitons and Reversible Manipulation on the Surface of a Prototype Obstructed Atomic Insulator, Silicon

Z. K. Liu[1*], P. Deng[2,3*], Y. F. Xu[4*], H. F. Yang[1], D. Pei[1,5], C. Chen[5,6], S. M. He[5], D. F. Liu[1], S. K. Mo[6], T. K. Kim[7], C. Cacho[7], H. Yao[8], Z. D. Song[4,9], X. Chen[2], Z. Wang[8], B. H. Yan[10], L. X. Yang[2], A. Bernevig[4,11,12†] and Y. L. Chen[1,5†]

[1]*School of Physical Science and Technology, ShanghaiTech Laboratory for Topological Physics, ShanghaiTech University, Shanghai 201210, China*
[2]*State Key Laboratory of Low Dimensional Quantum Physics, Department of Physics, Tsinghua University, Beijing, 100084, China*
[3]*Beijing Academy of Quantum Information Sciences, Beijing, 100193, China*
[4]*Department of Physics, Princeton University, Princeton, NJ, USA*
[5]*Department of Physics, Clarendon Laboratory, University of Oxford, Parks Road, Oxford OX1 3PU, UK*
[6]*Advanced Light Source, Lawrence Berkeley National Laboratory, Berkeley, CA 94720, USA*
[7]*Diamond Light Source, Harwell Campus, Didcot, OX11 0DE, UK*
[8]*Institute for Advanced Study, Tsinghua University, Beijing, 100084, China*
[9]*International Center for Quantum Materials, School of Physics, Peking University, Beijing 100871, China*
[10]*Department of Condensed Matter Physics, Weizmann Institute of Science, Rehovot 7610001, Israel*
[11]*Donostia International Physics Center, P. Manuel de Lardizabal 4, 20018 Donostia-San Sebastian, Spain*
[12]*IKERBASQUE, Basque Foundation for Science, Bilbao, Spain*

**Topologically trivial insulators can be classified into atomic insulators (AIs) and obstructed atomic insulators (OAIs) depending on whether the Wannier charge centers are localized or not at spatial positions occupied by atoms. An OAI can possess unusual properties such as surface states along certain crystalline surfaces, which advantageously appear in materials with much larger bulk energy gap than topological insulators, making them more attractive for potential applications. In this work, we show that a well-known crystal, silicon (Si) is a model OAI, which naturally explains some of Si's unusual properties such as its famous (111) surface states. On this surface, using angle resolved photoemission spectroscopy (ARPES), we reveal sharp quasi-1D massive Dirac line dispersions; we also observe, using scanning tunneling microscopy/spectroscopy (STM/STS), topological solitons at the interface of the two atomic chains. Remarkably, we show that the different chain domains can be reversibly switched at the nanometer scale, suggesting the application potential in ultra-high density storage devices.**

## I. Introduction

The classification of materials plays important roles in both physics and materials science: it shows the essential similarity of materials in a same class and the most significant difference between classes. In the past decade, the topological classification [1, 3-9] of compounds achieved great adoption, and led to the discovery of series of topological quantum materials including topological insulators and superconductors [10-14], topological Dirac and Weyl semimetals [15-19], and materials with fragile topology [20-23].

Despite the increasing number of discovered topological quantum materials, recent high throughput computations [1, 4-7, 24] show that the majority of all known materials are topologically trivial at the Fermi level. Thus, a question naturally arises: Do all the topologically trivial materials behave identically or are there clear distinct classes of behavior? Topologically trivial insulators, for example, can be classified into two types [2, 4, 25, 26] (see Fig. 1a): atomic insulators (AIs) and obstructed atomic insulators (OAIs), depending on whether the Wannier charge centers are localized or not at spatial positions occupied by atoms. Interestingly, an OAI can possess unusual properties such as surface states (SSs) along certain crystalline surfaces. Although required by the bulk OAI nature, these SSs do not necessarily cross the bulk gap, as they would do in a TI. However, they advantageously appear in materials with much larger bulk energy gap than TIs (the latter of which typically is smaller than ~0.3 eV), which makes them more attractive for potential applications. According to recent high-throughput computations [1, 2, 24-26], about 75% of all known insulators are topologically trivial at the Fermi level; 10% are OAIs and 65% are AIs [8] (see Fig. 1b).

To determine if an insulator is an OAI, one can compute its real space invariants (RSIs) [2, 21], which were developed under the framework of topological quantum chemistry [9, 27]. For a topologically trivial insulator, any non-zero-integer RSI index at an atomically empty Wyckoff position (WP) indicates an OAI [2, 21, 25]; the related empty WP is referred to as the obstructed Wannier charge center (OWCC). Under open-boundary conditions, a finite-size crystal of OAI with its OWCC on the surface (or edge/corner) presents the obstructed surface states (OSSs); the OSSs can be metallic or topological at the charge neutrality point (filling anomaly [28]) if the finite-size crystal preserves bulk crystalline symmetry.

## II. Results

### A. Verification of the OAI nature of Silicon

In the following, we explain the OAI nature of silicon (Si) using RSIs, which naturally predicts the existence of Si's famous (111) SSs [29]. Remarkbly, these SSs can persist under surface reconstructions as they do not affect the bulk OAI nature (see SI and discussions below for details). Si crystal belongs to the space group $Fd\bar{3}m$ (No. 227) and only the WP *8a* is occupied by Si atoms. In Ref. [2], the spinful band structure with the spin-orbit coupling (SOC) of Si is diagnosed as an OAI by the nonzero double-valued RSI at the empty WP *16d*. As the SOC of Si is so small that it does not change the topology of the occupied bands, in the present work, we have ignored its effect in the theoretical calculations and discussions. In the absence of SOC, the OAI phase of Si is indicated by the single-valued RSI $\delta_6(d) = 1$ (which is defined as: $\delta_6(d) = 3N(\Gamma_2^+) - N(\Gamma_3^-) - N(\Gamma_4^-) + 2N(\Gamma_5^-) - N(L_1^-) - 2N(L_2^+) + N(L_3^+)$, where $N(R)$ represents the multiplicity of the irreducible representation $R$ of the Bloch bands below the Fermi level ($E_F$), see SI for details). The nonzero RSI defined on WP *16d* confirms Si's OAI nature and the fact that any Miller plane cutting through WP *16d* – the (111) plane is one such plane (see Fig. 1c) – will give rise to OSSs in the bulk gap, which are the famous Si (111) SSs originally observed in 1962 [29]. Remarkably, as long as the bulk OAI phase of Si is unchanged and the position of OWCC is cleaved on the surface, the existence of OSSs is guaranteed, although surface modifications (e.g., surface reconstructions) may affect certain details of the OSSs.

We now further consider the surface reconstruction. The 2×1 surface reconstruction on Si (111) surface does not affect the existence of the OSSs, but does introduce band folding and in-gap dispersions at high symmetry points in the surface Brillouin zone (BZ) (see SI for details). Due to the reconstruction, atoms on the topmost layer of Si (111) surface form 1D chains with two types of buckling [30]: the negatively buckling chains (NBCs) and positively buckling chains (PBCs), respectively. By performing the slab calculations (see SI for details), we find that the 2D RSIs of the two types of slab structures, whose surfaces are formed by the respective NBCs and PBCs, are both nonzero but different (as detailed in the SI, the single-valued RSI for NBCs is {δ(*a*) = -7; δ(*b*) = -6} and for PBCs is: {δ(*a*) = -6; δ(*b*) = -7}[1], where *a* an *b* are the two WPs of the plane group (*Pm*) on the surface formed by NBC or PBC, respectively), consistent with Si's OAI nature. Further, the different RSIs for NBCs and PBCs suggest the valence states of NBCs and PBCs are in different atomic insulator phases, which cannot continuously evolve into each other without gap closing. This leads to the intriguing situation where, at the boundary between the 1D chains in the two reconstructed surfaces NBCs and PBCs of the SI obstructed atomic insulator, there must exist zero dimensional (0D) in-gap states, or solitons. These 0D solitons can further align to form 1D soliton lines (as can be seen in our experiments below), which become the boundary between the surface domains of NBCs and PBCs. As these (effectively) spinless solitons can carry the charge of one electron, they form interesting subjects for the study of low dimensional physics, such as the spin-charge separation [31-33] that was initially proposed in polyacetylene.

To verify the RSI analysis above and investigate the unique properties of OSSs on Si (111) surface in detail, we first construct a 1D atomic chain model to capture the overall characteristics including the 1D massive Dirac dispersions along the NBCs and PBCs, and soliton solutions at their interface; we then carry out *ab initio* calculations to obtain the detailed electronic structure.

As the interchain coupling between NBC and PBC chains is weak, similar to the 1D SSH [33] or polyacetylene [34] models, we can express the Hamiltonian for the NBC and PBC chains as (see SI for more details): $H_k = \epsilon_0 I_{(2X2)} - \hbar v_F k'_x \tau_x + m\tau_z$ where $\epsilon_0$ is the onsite energy, $\tau_x$ and $\tau_z$ are Pauli matrices, and *m* is the mass term. Because the two inequivalent Si atoms in the unit cell tilt oppositely for NBCs and PBCs, the mass terms for NBCs and PBCs have opposite sign, giving soliton solutions at the joint of NBCs and PBCs (see SI). This model analysis is in nice agreement with our *ab initio* calculations (see SI for details) which provide complete details of the 1D massive Dirac-line dispersions for NBCs/PBCs and the interface solitons (Fig. 1e).

## B. Measurement of the OSSs and Solitons in Silicon

In order to observe and manipulate the unique OSSs, we make use of state-of-the-art ARPES and STM/STS. By ARPES, we observe sharp 1D massive Dirac-line dispersions with weak inter-chain coupling; using STM/STS, we clearly resolve solitons at the interface between NBCs and PBCs, and demonstrate reversible switching of nanometer-sized NBC and PBC domains, suggesting the application potential in ultra-high density storage devices.

In our experiments, the Si (111) surface was obtained by cleaving a Si single crystal in-situ prior to ARPES/STM/STS measurements. The surface 1D atomic chains due to the 2×1 reconstruction can be clearly seen by STM topography mapping (Fig. 2a(i)), consistent with previous studies [35-39]. Due to the existence of three equivalent directions ($[0\bar{1}1]$, $[\bar{1}01]$ and $[1\bar{1}0]$) on the unreconstructed surface, the atomic chains after reconstruction can extend along three directions 120° apart, and Fig. 2a(ii) shows two such domains. Besides STM, the existence of different chain domains can also be verified by ARPES. By scanning the photon beam across the cleaved Si (111) surface, we observed band structures from one, two or all three differently oriented domains. As examples, Fig. 2b shows data from a region with one dominating domain whose Fermi-surface (FS) is comprised of one pair of parallel lines (as we'll discuss in detail below); while in Fig. 2c(i), FSs from all three differently oriented domains can be seen, forming a Kagome shaped network (more discussions can be found in SI).

The parallel line shaped FS (see Fig. 2b(i)) originates from the sharp 1D massive Dirac dispersions (see Fig. 2b(ii)), as predicted by the theoretical analysis above. Interestingly, the large distance between adjacent atomic chains (~0.65 nm, see Fig. 2a(i)), can introduce weak inter-chain coupling, which is evidenced by the slight decrease of Dirac velocity from the BZ center ($hv = 3.3$ eV·Å or $5\times10^5$ m/s, left panel of Fig. 2b(ii)) to the boundary ($hv = 2.8$ eV·Å or $4.25\times10^5$ m/s, right panel of Fig. 2b(ii)). The full evolution of the dispersions across the whole surface BZ is plotted in Fig. 2b(iii) and Fig. 2c(iii), clearly showing 1D Dirac lines across the whole BZ and agrees with our *ab initio* calculations (see SI). Finally, the surface nature of the sharp Dirac dispersions can be verified by photon energy dependent measurements [14] as illustrated in Fig. 2b(iv), where the straight vertical FS shows no $k_z$ dispersion.

In addition to the valance branch of the OSSs, the conduction branch can be seen in n-doped Si samples (Fig. 3a), whose 1D characteristic is again shown by parallel line shaped FSs (see Fig. 3a(i)) and the respective flat dispersions of narrow bandwidth (~30 meV, see Fig. 3a(iii, iv)). The mass term in the massive Dirac dispersion can now be estimated as ~150 meV at the BZ center (Fig. 3a(ii)) for NBCs, comparable to the calculated value (~100 meV). As the band velocity is different for NBCs and PBCs (due to their weak coupling to the underlying bulk atoms), by changing measurement positions on the sample surface, we were able to find dispersions from both NBC and PBC domains, with their velocity as ~3.3 eV·Å and ~2.8 eV·Å, respectively (see Fig. 3b), agreeing well with our *ab initio* calculations (more results and summary can be found in SI).

Besides their band dispersions, NBC and PBC domains can also be discerned by STM, allowing us to further search for their boundary and interface solitons. In Fig. 4a(i), a surface region showing both NBC and PBC domains is illustrated. By performing STS measurements along a chain crossing the NBC and PBC interface (marked by the white solid line in Fig. 4a(ii)), the spatial evolution of the STS spectra can be seen in Fig. 4a(iii), which shows the NBC/PBC type of STS spectra on the left/right, and strong in-gap soliton states in between, as clearly resolved for the first time (more details of the soliton spectra and analysis can be found in SI). In Fig. 4a(iv), three representative STS spectra from the NBC, PBC and soliton regions are plotted side-by-side for comparison, clearly showing their difference. Due to the presence of site-mirror reflection symmetry of the chains, solitons between NBCs and PBCs can exhibit intriguing phenomena such as spin-charge separation (when occupied, a soliton carries the charge of an electron while being spinless) along a soliton wire; and the lifted degeneracy between the NBCs and PBCs also provide an opportunity for bipolaron formation, which carry charge $-2e$ like a local Cooper pair, and a dilute density of charge $-2e$ bipolarons may result in strong coupling superconductivity [33, 40].

## C. Reversible Manipulation of the Solitons in Silicon

Interestingly, if external stimuli are applied, boundary solitons and the associated NBC/PBC domains can be manipulated. As an example, Fig. 4b shows two topography images of a same area before and after a sudden electric pulse discharged from the STM tip (see SI for details). Clearly, a segment of the domain boundary (green dashed line in Fig. 4b(ii)) shifts to the left, accompanied by an increase of the PBC domain size on the right. Such motion of the boundary soliton wires could be induced by the strong coupling between STM injected electrons and the local lattice vibration, similar to the atomic chains on the Ge (001) surface [41].

To systematically investigate this effect and explore effective methods to manipulate NBC and PBC domains, we carried out two sets of controlled experiments by continuously varying the STM bias voltage and current, respectively. Remarkably, both methods turned out to be effective and reversible switching of NBC/PBC domains can be achieved. As shown in Fig. 4c, two new NBC patches (5nm and 10nm in size, respectively) can be sequentially created by increasing the (negative) bias voltage above a threshold (about -0.35 V, see Fig. 4c(i-iii) and SI for more details); while under a positive bias voltage (above 1.5 V, see SI for details), the two NBC patches created above can be switched back to PBC patches, as illustrated in Fig. 4c(iv-vi) (more examples can be found in SI).

## III. Discussion and Conclusion

The reversible switching of the PBC and NBC domains is exciting, and could inspire potential applications for ultra-high density storage device. For example, the NBC and PBC domains can represent the "0" and "1" states, respectively; and the information can be "read out" by measuring the local density of state, which can differ by almost an order of magnitude at a proper bias voltage (e. g. within -0.3 V~-0.5 V, see Fig. 4a(iv) and SI for more discussions) for the NBC and PBC domains. On the other hand, the "writing" operation can be achieved by switching the NBC/PBC domains using proper electric voltage or current pulses as discussed above (see Fig. 4c). As the NBC/PBC domains being manipulated can be as small as a few of nano meters in size (~5 nm as seen in Fig. 4c), if used as storage bits, they promise significant increase in planar density (~ 25 T bit/in$^2$) compared to commonly used magnetic storage device nowadays (~ 2T bit/in$^2$)[42].

The understanding of Si as an OAI gives an example to show how the classification of topologically trivial insulators can provide new insights to materials including Si, one of the most studied materials on earth. On the other hand, based on the backbone material of the modern semiconductor industry, new properties and application potentials of Si are likely to foster devices that are not only fabrication friendly, but also of broad impact.

## IV. Methods
### A. Angle-resolved photoemission spectroscopy

ARPES measurements were performed at beamline I05 of Diamond light source (DLS, Proposal # SI9967), UK. and beamline 10.0.1 of the Advanced Light Source (ALS) at Lawrence Berkeley National Laboratory, USA. Samples were measured under ultra-high vacuum below $1\times10^{-10}$ Torr at DLS and $3\times10^{-11}$ Torr at ALS. Data were collected by Scienta R4000 analyzers at 10 K sample temperature. The total convolved energy and angle resolutions were 20 meV and 0.2°, respectively. In order to obtain a fresh and clean surface, the samples were cleaved in situ.

### B. Scanning tunnelling microscopy/spectroscopy

STM/STS experiments were carried out in ultra-high vacuum (UHV) below $1\times10^{-10}$ torr. Single crystals were cleaved *in situ* at room temperature then transferred to a cryogenic stage kept at 4.2 K for STM/STS experiments. Ag coated PtIr tips were used for imaging and tunnelling spectroscopy. The *dI/dV* spectra were obtained using a lock-in amplifier at a frequency of 913 Hz.

### C. LDA calculations

Electronic structures were calculated by the density-functional theory (DFT) method which is implemented in the Vienna *ab initio* Simulation Package (VASP) [43]. The core electrons were represented by the projected augmented wave method [44]. The exchange-correlation was considered in the generalized gradient approximation (GGA) [45] and the SOC effect of Si was ignored self-consistently. The energy cut off was set to be 300 eV for the plane-wave basis. Experimental lattice parameters were used in the construction of a slab model with seven-unit-cell thick to simulate a surface. Positions of outermost four atomic layers were fully optimized to consider the surface atomic relaxation.

**ACKNOWLEDGEMENTS**

This research used resources of the Advanced Light Source, a U.S. DOE Office of Science User Facility under contract no. DE-AC02-05CH11231. Y.L.C. acknowledges the support from the Oxford-ShanghaiTech collaboration project and the Shanghai Municipal Science and Technology Major Project (grant 2018SHZDZX02). B.A.B. received support from the European Research Council (ERC) under the European Union's Horizon 2020 research and innovation programme (grant agreement no. 101020833). B.A.B. received further support from the DOE Grant No. DE-SC0016239, the Schmidt Fund for Innovative Research, Simons Investigator Grant No. 404513, the Packard Foundation, the Gordon and Betty Moore Foundation through Grant No. GBMF8685 towards the Princeton theory program, and a Guggenheim Fellowship from the John Simon

Guggenheim Memorial Foundation. Further support was provided by the NSF-MRSEC Grant No. DMR-2011750, ONR Grant No. N00014-20-1-2303, BSF Israel US foundation Grant No. 2018226, and the Princeton Global Network Funds.

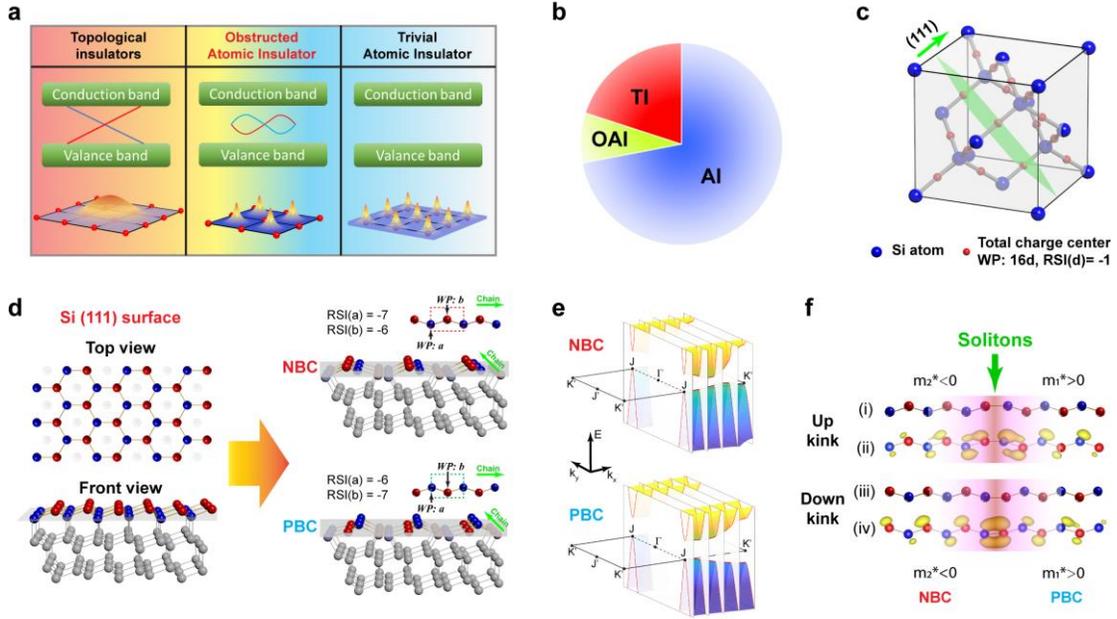

FIG. 1. Introduction to obstructed atomic insulator and Si (111) 2×1 surface states. (**a**) Illustration of three different insulating phases. Middle row: Characteristic electronic structures, showing existence of spin resolved topological surface states that necessarily connect the conduction and valence bands (left panel, spin polarization is marked by red and blue colors, respectively), obstructed surface states in the bulk gap that do not necessarily connect the conduction or valence bands (middle panel, can be spin resolved or not), and clean bulk gap without surface states (right panel). Bottom row: The distribution of Bloch functions in real space, showing non-localization (left panel), localization away from atoms (middle panel) and localization on atoms (right panel). (**b**) Pie chart showing portions of TIs, OAIs and AIs of all known insulators, estimated from high-throughput calculations [1, 2]. (**c**) Lattice structure of Si crystal (blue dots) and OWCCs (red dots) at the WP 16 *d*. Green plane indicates the (111) surface cutting through the OWCCs. (**d**) Schematic showing top layer atoms of Si (111) surface before (left panels) and after (right panels) the 2×1 reconstruction, forming NBC and PBC 1D chains. Both NBCs and PBCs are of plane group *Pm* with a vertical mirror plane perpendicular to the chain. The Wyckoff positions occupied by Si atoms are indicated along the chain. Green arrows on the right panels indicate the chains' direction. (**e**) Quasi-1D massive Dirac dispersions for NBC (top panel) and PBC (bottom panel) chains. Letters Γ, J, K', J' mark the high symmetry points of the surface BZ. (**f**) Formation of solitons at the intersections between NBC and PBC chains for both up and down kinks. Calculated electron density is indicated by the electron cloud (yellow) superimposed on chain atoms in (ii) and (iv), respectively. Acronyms: TI, topological insulator; OAI, obstructed atomic insulator; AI, trivial atomic insulator; OWCC, obstructed Wannier charge center; PBC, positively buckled chain; NBC, negatively buckled chain; WP, Wyckoff positions; RSI, real space invariant; BZ, Brillouin zone.

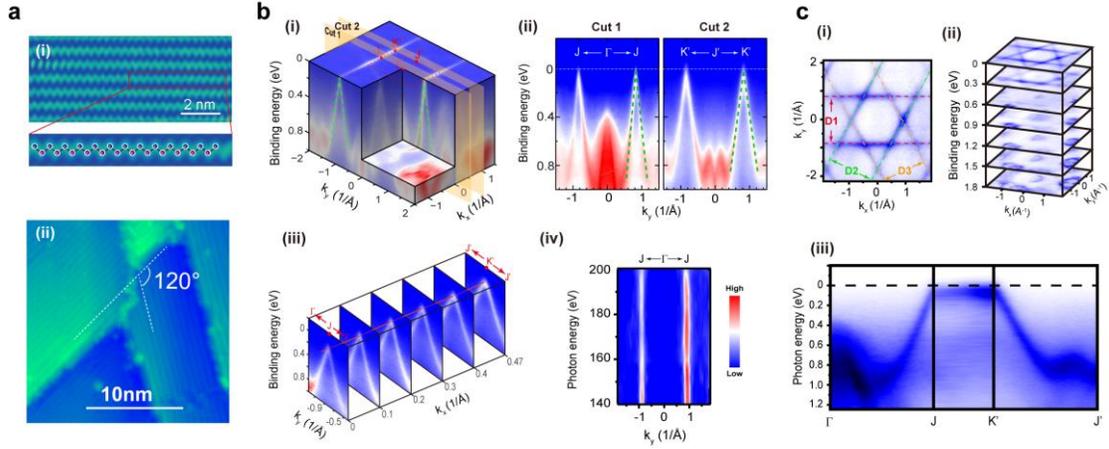

FIG. 2. 1D Dirac line band structure on Si (111) 2×1 surface. (**a**) STM topography showing 1D atomic chains on 2×1 reconstructed Si (111) surface with atom positions overlaid in (i) and two domains with chain direction apart by 120° in (ii). (**b**) (i) Three dimensional (3D) electronic structure plot shows a pair of 1D parallel line shaped Fermi-surface (FS) formed by sharp linear dispersions (marked by green dashed lines). The surface BZ is superimposed with high symmetry points labelled. (ii) Comparison of dispersions (see text for more details) along high symmetry directions of $J - \Gamma - J$ and $K' - J' - K'$, as indicated in panel (i). (iii) Six dispersion plots from the BZ center ($\Gamma - J - \Gamma$) to boundary ($J' - K' - J'$) show gradual evolution of the massive quasi-1D Dirac dispersions. (iv) $k_y$-$k_z$ FS map from photon energy dependent ARPES measurements shows vertical lines without $k_z$ dispersion. (**c**) (i) FS showing three parallel line pairs (D1, D2 and D3) from three domains with the respective chain directions apart by 120°. (ii) Stacked constant energy contour plots shows the energy dependent electronic structures. (iii) Surface state dispersions along high symmetry directions across the whole surface BZ.

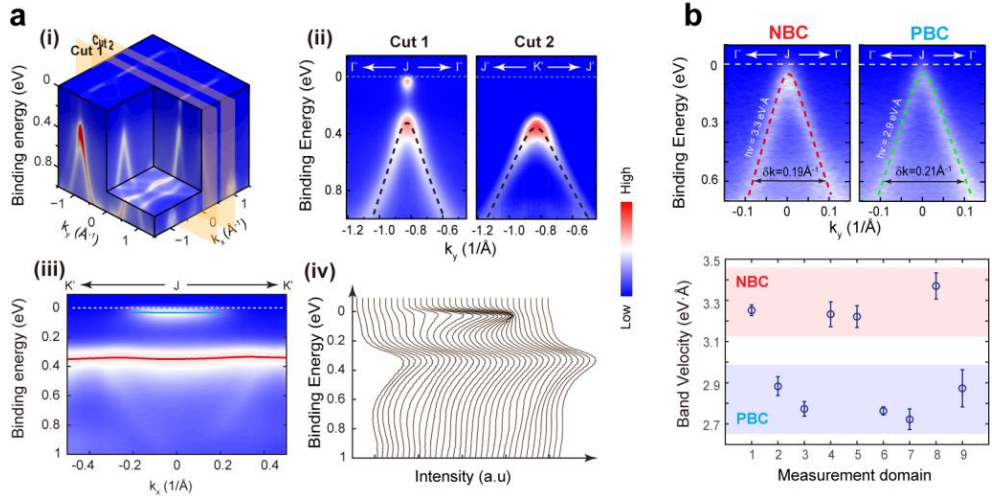

FIG. 3. Massive Dirac fermion and comparison between PBC and NBC. **(a)** (i) 3D plot of the band structure from an n-type sample, showing the quasi-1D shaped FS of the conduction bands of the OSSs, similar to that of the valance bands in Fig. 2B. (ii) Comparison of dispersions along high symmetry directions of $\Gamma - J - \Gamma$ and $J' - K' - J'$, as indicated in panel (i). A gap of ~300 meV (hence a mass term of 150 meV) can be estimated from the $\Gamma - J - \Gamma$ dispersion (Cut 1, left panel). (iii) Intensity plot and (iv) stacked line plot of the energy distribution curves along the $K' - J - K'$ direction extracted from (i). Overlapped blue and red curves in (iii) indicate the fitted dispersions of the surface conduction and valance bands, respectively. **(b)** Top panels: dispersions (along the $\Gamma - J - \Gamma$ directions) from PBC and NBC domains of a same sample, with the extracted dispersions (green and red broken lines) and respective linear band velocity labelled. (ii) Summary of the linear band velocity from nine measurements on different positions shows values centered around 3.3 eV·Å and 2.8 eV·Å for NBC and PBC domains, respectively.

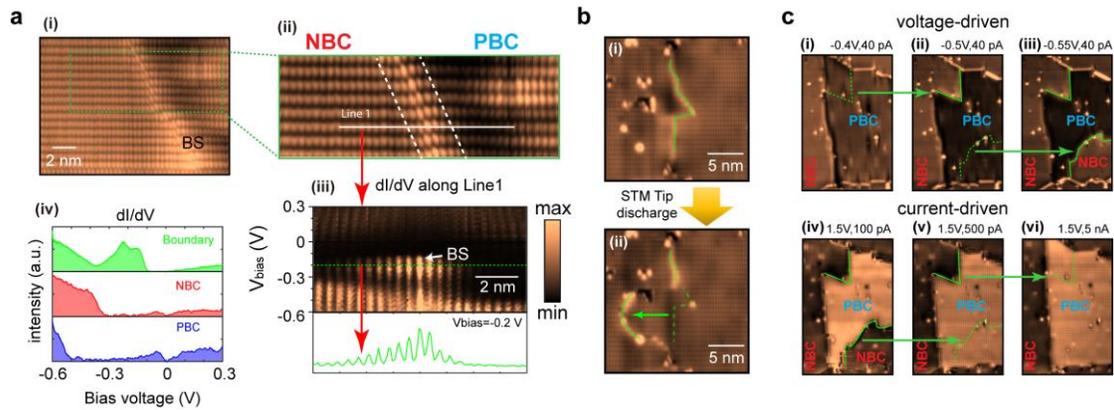

FIG. 4. Solitons and manipulation of PBC and NBC domains. **(a)** (i)-(ii) STM topography with atomic resolution (i), and a zoomed-in plot (ii) showing a clear boundary separating the NBC and PBC domains. (iii) dI/dV spectra along an atomic chain (indicated by line 1 in (ii)) crossing the NBC/PBC boundary. The green curve plots the evolution of the dI/dV intensity at $V_{bias}$=-0.2 V. BS: boundary state. (iv) Three representative dI/dV spectra from the boundary (green), PBC (blue) and NBC (red) chains, showing clear difference. **(b)** STM topography images of the same area before (i) and after (ii) a sudden electric pulse discharged from the STM tip. Green solid lines show the domain boundary between the NBC and PBC domains, and the green dashed line indicates the shifted domain boundary. **(c)** Sequential STM topography images from the same sample area showing the creation of two NBC patches (panel (i-iii), marked by the green arrows) by increasing negative STM bias voltage; and the back-switching of the two NBC patches to PBC type (panel (iv-vi), marked by the green arrows) by increasing the STM tip current under a positive bias voltage. The boundary of the NBC and PBC domains are tracked by green solid and dashed lines.

**References:**

[1] B. Bradlyn, L. Elcoro, J. Cano, M. G. Vergniory, Z. J. Wang, C. Felser, M. I. Aroyo, and B. A. Bernevig, *Topological quantum chemistry*, Nature **547**, 298-305 (2017).

[2] T. T. Zhang, Y. Jiang, Z. D. Song, H. Huang, Y. Q. He, Z. Fang, H. M. Weng, and C. Fang, *Catalogue of topological electronic materials*, Nature **566**, 475-479 (2019).

[3] M. G. Vergniory, L. Elcoro, C. Felser, N. Regnault, B. A. Bernevig, and Z. J. Wang, *A complete catalogue of high-quality topological materials*, Nature **566**, 480-485 (2019).

[4] F. Tang, H. C. Po, A. Vishwanath, and X. G. Wan, *Comprehensive search for topological materials using symmetry indicators*, Nature **566**, 486-489 (2019).

[5] H. C. Po, A. Vishwanath, and H. Watanabe, *Symmetry-based indicators of band topology in the 230 space groups*, Nature Communications **8**: 50 (2017).

[6] Y. F. Xu, L. Elcoro, Z.-D. Song, B. J. Wieder, M. G. Vergniory, N. Regnault, Y. L. Chen, C. Felser, and B. A. Bernevig, *High-throughput calculations of magnetic topological materials*, Nature **586**, 702-707 (2020).

[7] M. G. Vergniory, B. J. Wieder, L. Elcoro, S. S. P. Parkin, C. Felser, B. A. Bernevig, and N. Regnault, *All Topological Bands of All Stoichiometric Materials*, Science **376**, eabg9094 (2022).

[8] L. Elcoro, B. J. Wieder, Z. D. Song, Y. F. Xu, B. Bradlyn, and B. A. Bernevig, *Magnetic topological quantum chemistry*, Nature Communications **12**: 5965 (2021).

[9] Y. F. Xu, L. Elcoro, Z. D. Song, M. G. Vergniory, C. Felser, S. S. P. Parkin, N. Regnault, J. L. Mañes, and B. A. Bernevig, *Filling-Enforced Obstructed Atomic Insulators*, arXiv: 2106.10276 (2021).

[10] Y. F. Xu, L. Elcoro, G. W. Li, Z. D. Song, N. Regnault, Q. Yang, Y. Sun, S. S. P. Parkin, C. Felser, and B. A. Bernevig, *Three-Dimensional Real Space Invariants, Obstructed Atomic Insulators and A New Principle for Active Catalytic Sites*, arXiv:2111.02433 (2021).

[11] F. Schindler and B. A. Bernevig, *Noncompact atomic insulators*, Phys. Rev. B. **104**: L201114 (2021).

[12] C-K. Chiu, J. C. Y. Teo, A. P. Schnyder, and S. Ryu, *Classification of topological quantum matter with symmetries*, Reviews of Modern Physics **88**, 035005 (2016).

[13] J. Kruthoff, J. de Boer, J. van Wezel, C. L. Kane, and R-J. Slager, *Topological Classification of Crystalline Insulators through Band Structure Combinatorics*, Physical Review X **7**, 041069 (2017).

[14] X.L. Qi, and S.-C. Zhang, *Topological insulators and superconductors*, Reviews of Modern Physics **83**, 1057-1110 (2011).

[15] Y. Ando, and L. Fu, *Topological Crystalline Insulators and Topological Superconductors: From Concepts to Materials*, Annual Review of Condensed Matter Physics **6**, 361-381 (2015).

[16] B. A. Bernevig, T. L. Hughes, and S.-C. Zhang, *Quantum Spin Hall Effect and Topological Phase Transition in HgTe Quantum Wells*, Science **314**, 1757-1761 (2006).

[17] Y. L. Chen, J. G. Analytis, J. H. Chu, Z. K. Liu, S. K. Mo, X. L. Qi, H. J. Zhang, D. H. Lu, X. Dai, Z. Fang, S. C. Zhang, I. R. Fisher, Z. Hussain, and Z. X. Shen, *Experimental Realization of a Three-Dimensional Topological Insulator, $Bi_2Te_3$*, Science **325**, 178-181 (2009).

[18] X. G. Wan, A. M. Turner, A. Vishwanath, and S. Y. Savrasov, *Topological semimetal and Fermi-arc surface states in the electronic structure of pyrochlore iridates*, Physical Review B **83**, 205101 (2011).

[19] Z. K. Liu, B. Zhou, Y. Zhang, Z. J. Wang, H. M. Weng, D. Prabhakaran, S. K. Mo, Z. X. Shen, Z. Fang, Z. Fang, X. Dai, Z. Hussain, and Y. L. Chen, *Discovery of a Three-Dimensional Topological Dirac Semimetal, $Na_3Bi$*, Science **343**, 864-867 (2014).

[20] B. Q. Lv, H. M. Weng, B. B. Fu, X. P. Wang, H. Miao, J. Ma, P. Richard, X. C. Huang, L. X. Zhao, G. F. Chen, Z. Fang, X. Dai, T. Qian, and H. Ding, *Experimental Discovery of Weyl Semimetal TaAs*, Physical Review X **5**, 031013 (2015).

[21] D. F. Liu, A. J. Liang, E. K. Liu, Q. N. Xu, Y. W. Li, C. Chen, D. Pei, W. J. Shi, S. K. Mo, P. Dudin, T. Kim, C. Cacho, G. Li, Y. Sun, L. X. Yang, Z. K. Liu, S. S. P. Parkin, C. Felser, and Y. L. Chen, *Magnetic Weyl semimetal phase in a Kagomé crystal*, Science **365**, 1282-1285 (2019).

[22] B. H. Yan, and C. Felser, *Topological Materials: Weyl Semimetals*, Annual Review of Condensed Matter Physics **8**, 337-354 (2017).

[23] H. C. Po, H. Watanabe, and A. Vishwanath, *Fragile Topology and Wannier Obstructions*, Physical Review Letters **121**, 126402 (2018).

[24] V. Peri, Z. D. Song, M. Seera-Garcia, P. Engeler, R. Queiroz, X. Q. Huang, W. Y. Deng, Z. Y. Liu, B. A. Bernevig, and D. H. Sebastian, *Experimental characterization of fragile topology in an acoustic metamaterial*, Science **367**, 797-800 (2020).

[25] J. Cano, B. Bradlyn, Z. J. Wang, L. Elcoro, M. G. Vergniory, C. Felser, M. I. Aroyo, and B. A. Bernevig, *Building blocks of topological quantum chemistry: Elementary band representations*, Physical Review B **97**, 035139 (2018).

[26] Z. D. Song, L. Elcoro, Y. F. Xu, N. Regnault, and B. A. Bernevig, *Fragile Phases as Affine Monoids: Classification and Material Examples*, Physical Review X **10**, 031001 (2020).

[27] H. Watanabe, H. C. Po, and A. Vishwanath, *Structure and topology of band structures in the 1651 magnetic space groups*, Science Advances **4**: eaat8685 (2018).

[28] W. A. Benalcazar, T. Li, and T.L. Hughes, *Quantization of fractional corner charge in $C_n$-symmetric higher-order topological crystalline insulators*, Physical Review B **99**, 245151 (2019).

[29] F. G. Allen, and G.W. Gobeli, *Work Function, Photoelectric Threshold, and Surface States of Atomically Clean Silicon*, Physical Review **127**, 150-158 (1962).

[30] F. Ancilotto, W. Andreoni, A. Selloni, R. Car, and M. Parrinello, *Structural, electronic, and vibrational properties of Si(111)-2x1 from ab initio molecular dynamics*, Physical Review Letters **65**, 3148-3151 (1990).

[31] W. P. Su, J. R. Schrieffer, and A. J. Heeger, *Solitons in Polyacetylene*, Physical Review Letters **42**, 1698-1701 (1979).

[32] R. Jackiw, and C. Rebbi, *Solitons with fermion number 1/2*, Physical Review D **13**, 3398-3409 (1976).

[33] A. J. Heeger, S. Kivelson, J. R. Schrieffer, W. P. Su, and A. J. Heeger, *Solitons in conducting polymers*, Reviews of Modern Physics **60**, 781-850 (1988).

[34] W. P. Su, J. R. Schrieffer, and A. J. Heeger, *Soliton excitations in polyacetylene*, Physical Review B **22**, 2099-2111 (1980).

[35] K. C. Pandey, *New π-Bonded Chain Model for Si(111)-(2x1) Surface*, Physical Review Letters **47**, 1913-1917 (1981).

[36] J. E. Northrup, and M.L. Cohen, *Reconstruction Mechanism and Surface-State Dispersion for Si(111)-(2x1)*, Physical Review Letters **49**, 1349-1352 (1982).

[37] G. Xu, B. C. Deng, Z. X. Yu, S. Y. Tong, M. A. van Hove, F. Jona, and I. Zasada, *Atomic structure of the cleaved Si (111)-2x1 surface refined by dynamical LEED*, Physical Review B **70**, 045307(2004).

[38] R. I. G. Uhrberg, G. V. Hansson, J. M. Nicholls, and S. Flodström, *Experimental Evidence for One Highly Dispersive Dangling-Bond Band on Si(111) 2x1*, Physical Review Letters **48**, 1032-1035 (1982).

[39] P. Mrtensson, A. Cricenti, and G.V. Hansson, *Photoemission study of the antibonding surface-state band on Si(111)2x1*, Physical Review B **32**, 6959-6961 (1985).

[40] A. Alexandrov, and J. Ranninger, *Bipolaronic superconductivity*, Physical Review B **24**, 1164-1169 (1981).

[41] K. Tomatsu, K. Nakatsuji, M. Yamada, F. Komori, B. H. Yan, C. Y. Yam, T. Frauenheim, Y. Xu, and W. H. Duan, *Local Vibrational Excitation through Extended Electronic States at a Germanium Surface*, Physical Review Letters **103**, 266102 (2009).

[42] E. Grochowski, and P. Goglia, *The Magnetic Hard Disk Drive-Today's Technical Status And Future*, in SNIA Data Storage (DSC) Conference. 2016.

[43] G. Kresse, and J. Furthmüller, *Efficient iterative schemes for ab initio total-energy calculations using a plane-wave basis set*, Physical Review B **54**, 11169-11186 (1996).

[44] P. E. Blöchl, *Projector augmented-wave method*, Physical Review B **50**, 17953-17979 (1994).

[45] J. P. Perdew, K. Burke, and M. Ernzerhof, *Generalized Gradient Approximation Made Simple*, Physical Review Letters **77**, 3865-3868 (1996).